\documentclass[english]{article}
\usepackage{mathptmx}
\usepackage[T1]{fontenc}
\usepackage[latin9]{inputenc}
\usepackage{geometry}
\usepackage{amsmath}
\usepackage{amssymb}
\usepackage{hyperref}
\usepackage{tikz,pgf}
\usepackage[authoryear]{natbib}

\makeatletter

\usepackage{soul}

\AtBeginDocument{
  
}

\makeatother

\usepackage{babel}
\begin{document}
\title{Comment on "Blessings of Multiple Causes''}
\author {Elizabeth L. Ogburn, Ilya Shpitser, and Eric J. Tchetgen Tchetgen}

\maketitle

\emph{This document was updated on October 16, 2019 to include a brief rejoinder to WB's response to the body of our comment.  Their response can be found \href{https://arxiv.org/abs/1910.07320}{here}; our rejoinder is in the final section of this comment.  We are grateful to WB for their response, and especially for pointing out some errors and points of clarification in a previous version of this comment.}

\vspace{.5cm}

We are grateful to Wang and Blei (2019) (hereafter WB) for drawing attention to the important and increasingly popular project of using latent variable methods to control for unmeasured confounding. Prior causal inference research on this topic has not been adequately communicated or disseminated, leaving room for misconceptions, which we hope to begin to remedy in this discussion.  We also appreciate that the authors have sought and been receptive to our feedback about their work.  We would also like to thank the editors for giving us the opportunity to comment on this paper.

However, this paper has foundational errors.  Specifically, the premise of the \emph{deconfounder}, namely that a variable that renders multiple causes conditionally independent also controls for unmeasured multi-cause confounding, is incorrect. This can be seen by noting that no fact about the observed data alone can be informative about ignorability, since ignorability is compatible with any observed data distribution.  Methods to control for unmeasured confounding may be valid with additional assumptions in specific settings (e.g. \citealp{price2006principal, kuroki2014measurement, angrist2008mostly}), but they cannot, in general, provide a checkable approach to causal inference, and they do not, in general, require weaker assumptions than the assumptions that are commonly used for causal inference. While this is outside the scope of this comment, we note that much recent work on applying ideas from latent variable modeling to causal inference problems suffers from similar issues.



Causal inference aims to draw inferences about the parameters of the full data distribution--the distribution of the observed random variables and the potential outcomes--from 
realizations of the observed data distribution, which is generally a coarsened version of the full data distribution.  For example, the full data distribution for a conditionally ignorable model with binary treatment is of the form $p(Y(1), Y(0), A, {\bf X})$, where the following conditional independences hold on the counterfactual outcomes $Y(1)$, and $Y(0)$, the treatment $A$ and the set of baseline covariates ${\bf X}$: $Y(1) \perp A | {\bf X}$, and $Y(0) \perp A | {\bf X}$.   The parameter of interest is often the average causal effect (ACE): $\mathbb{E}[Y(1) - Y(0)]$.  The observed data distribution, on the other hand, is of the form $P(Y, A, {\bf X})$, where the observed outcome $Y$ is a coarsened version of $Y(1)$ and $Y(0)$, defined by consistency as $Y(1) \cdot A + Y(0) \cdot (1 - A)$.  Causal inference problems are often viewed as missing data problems, since every realization of the observed outcome $Y$ yields exactly one of the potential outcomes for the corresponding subject, with the other outcomes being missing data. With the deconfounder, WB aim to tackle 
settings with a vector ${\bf A}$ of multiple treatments, where 
baseline covariates are unobserved (except for single-cause confounders, which we ignore throughout). In such cases, the observed data distribution is a marginal distribution of the form $p(Y, {\bf A})$, marginalized over the missing potential outcomes and the unobserved confounders.

The deconfounder proposal can be loosely summarized as follows:
\begin{itemize}
\item Suppose \emph{ignorability} for the effect of a vector of causes $\mathbf{A}$ on an outcome $Y$ holds
conditional on $U$: $\mathbf{A}\perp Y(\mathbf{a})|U$.
\item $U$ is unobserved, but if it were observed 
then conditioning on and standardizing by $U$  (\emph{covariate adjustment}, or the \emph{adjustment formula}) would identify causal effects
of $\mathbf{A}$ on $Y$, as in equation (2) of WB.  
\item In lieu of the unmeasured $U$, and in the absence of any unmeasured
single-cause confounders, one can control for any variable $Z$ such
that $\text{A}_{1},...,A_{m}$ are mutually independent conditional
on $Z$, because such a $Z$ satisfies ingorability for all multi-cause confounders. $Z$ is a \emph{substitute confounder} for the true confounder $U$.
\end{itemize}

In addition to the above, the authors impose several additional assumptions at various points throughout the paper.  We describe these below. Nevertheless, the assumptions, as stated, do not imply
the claimed results.  

\section{Conditionally independent causes do not ensure conditional ignorability.}

The third step is the crux of the deconfounder. 
However, the criterion of conditional independence does not suffice to make $Z$ a valid substitute confounder.  This criterion does not rule out the inclusion of variables that may bias effects, nor does it ensure that all multi-cause confounders are captured by $Z$.
Finding an observed proxy that suffices to control for all confounding via covariate adjustment is related to a body of work on {\it complete adjustment criteria} \citep{shpitser10on,perkovic15complete}.  Below we give a few examples that violate these adjustment criteria, meaning that covariate adjustment is not a valid identification strategy, but that are not excluded from the deconfounder.

\subsection{The deconfounder may include variables that bias effects.}
 
\begin{figure}
\begin{centering}
\begin{tikzpicture}[>=stealth, node distance=0.9cm]
    \tikzstyle{format} = [ very thick, circle, minimum size=5.0mm,
	inner sep=0pt]
    \tikzstyle{square} = [very thick, rectangle, draw]

         \begin{scope}[xshift=0cm]
                \path[very thick, ->]
                        node[format] (a) {$A_1$}
                        node[format, right of=a] (r) {$R$}
                        node[format, above of=r, yshift=.3cm] (z) {$U$}
                        node[format,  right of=r] (b) {$A_2$}
                        node[format, right of=b] (c) {$A_3$}
                        node[format, right of=c] (y) {$Y$}
                        
                        node[format, right of=z] (v) {$V$}

                        (v) edge[dashed] (r)
                        (v) edge[dashed, bend left] (y)
                        
                        (z) edge (a)
                        (z) edge (b)
                        (z) edge (y) 
                        (a) edge (r)
                        (r) edge (b)
                        (r) edge [bend right] (y)
                        (r) edge [bend right] (c)
		node[below of=a, xshift=1cm, yshift=0.1cm] (l) {$(a)$}
                    
                ;
        \end{scope}
        \begin{scope}[xshift=6cm]
                \path[very thick, ->]
                        node[format] (a) {$A_1$}
                        node[format, right of=a] (d1) {$\ldots$}
		       node[format, right of=d1] (b) {$A_m$}
                        node[format, above of=d1, yshift=.5cm] (z) {$U$}
                        node[format, right of=b] (d) {$D$}
                        node[format, right of=d] (y) {$Y$}
                                  
                        (z) edge (a)
                        (z) edge (b)
                        (z) edge (y) 
                        (b) edge (d)
                        (d) edge (y)
                        (z) edge  [dashed,->]  (d)

		node[below of=a, xshift=1cm, yshift=0.1cm] (l) {$(b)$}
                    
                ;
        \end{scope}
	
	\begin{scope}[yshift=-3.5cm]
		\path[->, very thick]
		       node[format, yshift=.4cm] (a) {$A_1$}
                        node[format, right of=a, xshift=-.2cm] (d1) {$\ldots$}
		       node[format, right of=d1, xshift=-.2cm] (b) {$A_m$}
                        node[format, above of=b, yshift=.2cm] (z) {$U$}
                        node[format, right of=b, xshift=.7cm] (y) {$Y$}
                        node[format, below right of=b, yshift=0.1cm] (c) {$C$}
          
                        (z) edge (a)
                        (z) edge (b)
                        (z) edge (y)
                        (b) edge (c)
                        (y) edge (c)
                        (z) edge  [bend left=20, dashed,->]  (c)

			node[below of=a, xshift=1cm, yshift=-0.3cm] (l) {$(c)$}

			;
	\end{scope}
	\begin{scope}[xshift=6cm, yshift=-3.5cm]
		\path[->, very thick]
		       node[format] (a) {$A_1$}
                        node[format, right of=a] (d1) {$\ldots$}
		       node[format, right of=d1] (b) {$A_m$}
                        node[format, above of=b, yshift=.5cm] (z) {$U$}
                        node[format, below right of=z, xshift=.2cm] (m) {$M$}
                        node[format, above right of=m] (v) {$V$}
                        node[format, right of=b, xshift=.7cm] (y) {$Y$}
          
                        (z) edge (a)
                        (z) edge (b)
                        (z) edge (m)
                        (v) edge (m)
                        (v) edge (y)

			node[below of=d1, xshift=1cm, yshift=0.1cm] (l) {$(d)$}
			;
	\end{scope}

	\end{tikzpicture} 
\par\end{centering}
\caption{(a) A DAG in which $A_1$ and $A_2$ are causally dependent.  (b) A DAG with a single-cause mediator.
(c) A DAG with a single-cause collider. (d) A DAG with an M-bias collider.}
\label{fig:DAGs} 
\end{figure}
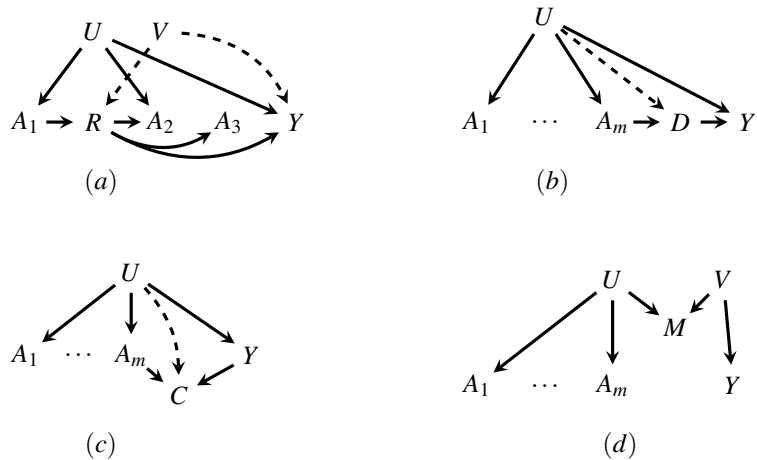

A substitute confounder constructed in order to render the causes mutually independent may include three types of variables that undermine
the ability to identify causal effects. M-bias colliders, such as
$M$ in the directed acylic graph (DAG) in Figure \ref{fig:DAGs} (d), and single-cause colliders, such as $C$ in Figure \ref{fig:DAGs} (c), are variables that \textit{induce} confounding \citep{cole2009illustrating, elwert2014endogenous}, and single-cause mediators, such as $R$ in Figure \ref{fig:DAGs} (a) and $D$ in Figure \ref{fig:DAGs} (b), are variables that bias causal effects.

Both colliders and mediators are post-treatment variables. As WB note, it is crucial that all covariates used to identify causal effects via the formula in (2) be pre-treatment
variables, because conditioning on a downstream effect of $\mathbf{A}$
may introduce bias in any direction (it need not bias effect estimates
towards the null). 
However, Lemma 4, which states that the substitute confounder $Z$ is guaranteed to be pre-treatment,
is incorrect. We first give an intuitive counterexample in which mediators would be included in a substitute confounder and then point out a problem in the proof of Lemma 4.

\paragraph{Causes cannot be causally dependent.}
Suppose the causes $A_{1},...,A_{m}$ can themselves have causal effects
on one another, as would be expected in most of the motivating examples
described in the introduction (neurons may cause one another
to fire; enrolling in one social program may increase the chance
that someone will learn of or be referred to another social
program; one medicine may be prescribed to treat side effects of another
or of a procedure). Specifically, consider the case depicted in Figure
\ref{fig:DAGs} (a), where $A_{1}$ causes $A_{2}$, and in order to render them conditionally
independent the deconfounder must include a variable, $R$, that breaks
this connection. However, the effect of $A_1$ on $Y$ is \emph{through} $R$ and therefore cannot be identified
controlling for $R$; depending on the relationships among $A_{1}$,
$R$, and $Y$, an estimator that controls for $R$ could either over-
or underestimate the true effect. This scenario is directly analogous to longitudinal causal inference problems with multiple time-varying treatments that contain time-varying confounders, variables that serve as confounders for some treatments and as mediators for other treatments. If there is an unmeasured confounder for the $R$-$Y$ relationship (represented by $V$ and the dashed arrows in Figure 1 (a)), then conditioning on $R$ fails to identify the direct effects of $\mathbf{A}$ on $Y$, because it opens a confounding pathway through $V$. See \cite{hernan2020causal} for an overview of these issues.

The answer to the question posed in Appendix B of WB, "Can the causes be causally dependent
among themselves?'' is therefore "no."  If they are causally dependent then the deconfounder, by dint of rendering the causes independent, breaks some of the structure among the causes  $\bf{A}$, and as was originally established in the time-varying treatment setting, this undermines the identification of joint effects of $\bf{A}$ on $Y$ by covariate adjustment.

\paragraph{Analysis of Lemma 4.}This simple argument also serves as a counterexample to Lemma 4, which states that the deconfounder does not pick up any post-treatment variables and can be treated as a pre-treatment covariate.
This is necessarily false whenever the causes are causally dependent among themselves, but it need
not hold even if the causes are not causally dependent, see below.

The proof of Lemma 4 in Appendix I states that "Inferring
the substitute confounder $Z_{i}$ is separated from estimating the
potential outcome. It implies that the substitute confounder is independent
of the potential outcomes conditional on the causes.''  The proof invokes the assumption that  $Z \perp Y({\bf A}) | {\bf A}$.  By the consistency property in causal inference, which defines the observed data variable $Y$ as $\sum_{{\bf a}} \mathbb{I}({\bf A}={\bf a}) Y({\bf a})$, $Y(\mathbf{A})$ is equal to $Y$, which implies $Z \perp Y | {\bf A}$.  This conditional independence cannot hold for any $Z$ that satisfies ignorability, except in trivial settings. Limiting inquiry to settings in which there exists a deterministic function of the causes that suffices to identify causal effects rules out almost everything that is typically considered confounding.\footnote{We updated the statement of this result to reflect the fact that different definitions exist for the presence of confounding; thank you to WB for drawing to our attention the fact a previous version was not entirely clear.}  
(This is also the case replacing $Y({\bf A})$ with $Y({\bf a})$ in the original assumption (note the lower case ${\bf a})$, since $Y$ is a fixed function of $Y({\bf a})$ and ${\bf A}$.)

In fact, confounders confound \textit{because} they
are related to potential outcomes even conditional on the observed
treatment and outcome. For example, if a person knows that their potential pain status under treatment $A=tylenol$ is 
preferable to their potential pain status under treatment $A=no$  $tylenol$, then they are more likely to take tylenol when they have a headache--so $Y(a)$ affects $A$.  Obviously $Y(a)$ also affects $Y$, their pain status after treatment, so $Y(a)$ is itself a confounder.  While this may be an extreme example, in general confounders are, almost by definition, intricately linked to the potential outcomes.


\vspace{.5cm}
When the causes are not causally dependent (which is the setting for which WB recommend using the deconfounder, see Appendix B), can we ensure that a substitute confounder does not contain post-treatment variables? 
Any mediator or collider caused by more than one cause will be excluded
from the substitute confounder, because such a variable  is a
collider between its causes and conditioning on it induces, rather
than eliminates, dependence among the causes.  But single-cause mediators and colliders may be incorporated into the substitute confounder.

\paragraph{Single-cause mediators and colliders.}
A single-cause mediator, such as $D$ in Figure \ref{fig:DAGs} (b) will generally not be required in order to render the causes conditionally independent, and the same is true of a single cause collider, such as $C$ in Figure \ref{fig:DAGs} (c).  But in the absence of Lemma 4 we do not immediately see any way to guarantee that single-cause
mediators and colliders would be excluded from substitute confounders.  In particular, if the dashed arrow in Figure \ref{fig:DAGs} (b) is present, so that the unmeasured confounder is not independent of the mediator, then it seems plausible that a substitute confounder would include some or partial information about the mediator.  Similarly, if the dashed arrow in Figure \ref{fig:DAGs} (c) is present, so that the unmeasured confounder is not independent of the collider, then it seems plausible that a substitute confounder would include some or partial information about the collider.

\paragraph{M-bias colliders.}
Even if one could exclude post-treatment random variables from the deconfounder, 
M-bias colliders, like $M$ in Figure 1 (d), can be pre-treatment.  They provide a counterexample to the premise
that a pre-treatment $Z$ that renders the causes conditionally independent suffices
to control for multi-cause confounding of A on Y, and specifically
to Lemmas 1 and 2.  While conditioning on $U$ itself would suffice to control for M-bias, if, in addition to $M$, $Z$ captures the part of $U$ that affects dependence among the causes without capturing the part of $U$ that relates $A_m$ to $M$, then M-bias would remain.\footnote{We are grateful to WB for catching a mistake in a previous version of this counterexample.}

\subsection{The deconfounder need not capture all multi-cause confounders.}


We provide an example to illustrate that the deconfounder may not capture all multi-cause confounders, and then we point out a flawed premise in the proof of Lemmas 1 and 2.  A related point is that the deconfounder may not be able to control for confounding even if it does capture all multi-cause confounders; this is because confounding involves the joint distribution of the causes and the potential outcomes, so in general learning a latent confounder requires dealing with this joint distribution.  This is established by a copula argument in \cite{d2019multi}.



Conditioning on $Z$ can render the causes mutually independent by separating a multi-cause confounder $U$ into single-cause components, while failing to control for the relationship between the causes and the outcome.  Here is an example: suppose $U$ is a confounder of $A_1$, $A_2$, and $Y$, and suppose that, conditional on $Z$, $U\sim Unif(0,1)$.  Then $U|Z$ is decomposable into the sum of $V$ and $W$, where $V$ and $W$ are independent.\footnote{A random variable is decomposable if it is equal to the sum of independent random variables; a $Unif(0,1)$ random variable is decomposable into a bernoulli random variable that takes the values 0 or .5 with equal probability and a uniform random variable over (0,.5).}  Further suppose that $A_1$ only depends on $V$ and $A_2$ only depends on $W$.  Then conditioning on $Z$ renders $A_1$ and $A_2$ independent, but there is no reason to think that it controls for confounding by $U$.

This counterexample to the claim that the deconfounder controls for all multi-cause confounding is pathological, but given the fact that modeling the marginal distribution of the causes can only tell us about the joint distribution of the causes and the outcome under stringent assumptions or in degenerate models, we expect counterexamples to be the rule, not the exception.

\paragraph{Analysis of Lemmas 1 and 2.}
The discussion above undermines the claim that the deconfounder, estimated via a factor model of the causes, suffices for ignorability to hold.  The argument for this claim in WB is rather technical, but we briefly analyze it here.  It is made through Lemma 1, which states that if $\bf{A}$ admits a \emph{Kallenberg construction} from the deconfounder then ignorability holds conditional on the deconfounder, and Lemma 2, which states that all factor models of $\bf{A}$ admit a Kallenberg construction.  However, Definition 3 misstates the Kallenberg construction for the relevant probability model.  The probability model for analyzing the causal effect of $\bf{A}$ on $Y$ subject to confounding by $Z$ is the model for the full data distribution; the probability model that includes only the observed data is appropriate for prediction but not for causal inference.  The full data comprise (in chronological/causal order) the random variables $\{Y(\mathbf{a}):\mathbf{a}\in\mathcal{A}\}$, $Z$, $\bf{A}$, and $Y$.  Note that ignorability is a restriction on the full data distribution, not the observed data distribution (which often has no restrictions in causal inference problems).  Put another way,
 no fact about the observed data alone can be informative about ignorability, since ignorability is compatible with any observed data distribution. 
Therefore, Theorem 5.10 of  \cite{kallenberg2006foundations} in fact implies $A_{ij}\overset{a.s.}{=}f_j(Z_i,\{Y(\mathbf{a}):\mathbf{a}\in\mathcal{A}\},U_{ij})$ rather than the construction given in equation (37) of WB, which omitted $\{Y(\mathbf{a}):\mathbf{a}\in\mathcal{A}\}$.
Thus, the Kallenberg construction used in the paper cannot link factor models to ignorability.  A Kallenberg construction on the full data, which could be informative about ignorability, is impossible to obtain given observed data information alone.


\subsection{When would a latent substitute confounder be expected to control for all multi-cause confounding?}

Identifying a latent substitute confounder from the observed data on $\mathbf{A}$ essentially requires
 the assumption that learning structure on the causes suffices to learn about any joint structure linking the causes with the outcome, in addition to the assumptions above.
 
A widely studied setting in which this would hold is when $U$ represents unknown structure that is common to each $A_k$ and to $Y$.
This is likely to be the case in GWAS studies and in problems with clustered data with unknown clusters.
In GWAS studies, including in WB's simulations, $U$ represents population structure that is common across all of the causes and the outcome.  For example, $U$ might be an ancestry matrix indicating how $n$ subjects are related to one another, and each of the causes and the outcome are expected to show dependence across the $n$ subjects due to this same ancestry matrix.  In this setting, any subset of the collection of variables with this same structure, that is any subset of $\{A_1,...,A_m,Y\}$, can be used to learn the common underlying population structure, in particular the set $\{A_1,...,A_m\}$ as is commonly done in practice \citep{price2006principal}.  

Theorem 6 requires the deconfounder to be piecewise constant in the causes; this reduces the problem of confounding to one of clustering.

Another example when a latent substitute confounder controls for all multi-cause confounding is the fully parametric model given in Appendix C of WB.

\section{Assumptions beyond ignorability}

In this section we assume that we are in the class of problems for which latent substitute confounders are known to perform well, 
e.g. in the GWAS or clustering setting.  We argue that even for those limited settings the assumptions required of the deconfounder are quite strong, and are not nonparametric.  Below we discuss the assumptions required for the deconfounder that go beyond those required for "classical causal inference."  In exchange for the assumptions listed below, "classical causal inference" requires the sole (but strong and untestable) assumption of no unmeasured multi-cause confounders.  Both the deconfounder and classical methods require no unmeasured single-cause confounders, SUTVA, and overlap (or positivity).

\paragraph{Nonparametric identification.}
Although the terms \emph{parametric} and \emph{nonparametric} can mean different things to different researchers, generally a causal effect is said to be \emph{nonparametrically identified} if either (a) the assumptions required for identification place no restrictions on the observed data distribution, except possibly up to a set of distributions of measure zero \citep{bickel1993efficient}, or (b) the only restrictions on the observed data distribution are those imposed by a nonparametric structural equation model. Such restrictions may include some conditional independences and inequality constraints. 
But causal effects cannot be nonparametrically identified (in either sense) in the setting considered in WB; identification requires assumptions that place substantial restrictions on the observed data distribution and on the structural equation models.

\paragraph{Semi-parametric and parametric assumptions.}
Contrary to its statement, Theorem 6, which identifies the joint causal effect of all of the causes on $Y$, rests on the parametric assumptions that the confounding variable is a clustering indicator and that the treatment effects are constant across clusters (no treatment-confounder interaction).  Furthermore, although it is not listed in the assumptions in the paper, in order for $f_1(\mathbf{a},x)$ and $f_2(z)$ to be jointly estimable even though $z$ is a deterministic function of $\mathbf{a}$, Theorem 6 also requires $f_1$ to be more smooth than $f_2$, e.g. they cannot be collinear.

Theorem 7 identifies the causal effect of a subset of $k$ out of the $m$ causes, assuming overlap/positivity for those $k$ causes: $P((A_1,...,A_k)\in\mathcal{A}|Z_i)>0$ for any set $\mathcal{A}$ such that $P(\mathcal{A})>0$.  Because the conditioning event $Z_i$ is a deterministic function of ${A_1,...,A_m}$, this is a stronger assumption than the classical overlap assumption, and it greatly restricts the possible functional forms for the deterministic function of $\bf{A}$ that gives $Z$.  This restriction will be greatest when $k$ is close to $m$.  Two open questions are (1) whether these restrictions imply that the model for $Z$ is degenerate as $m\rightarrow \infty$ and (2) whether they restrict the observed data distribution in addition to restricting the function of $\bf{A}$ that gives $Z$. Neither of these concerns is addressed in the paper, leaving open the possibility that the statement of the theorem might be vacuous, requiring parametric and/or additional causal assumptions in order for these conditions to be met.\footnote{We are grateful to WB for pointing out that a previous version of this statement was imprecise.} This framework, but with $k<<m$ and the addition of parametric assumptions and exclusion restrictions (i.e. that most causes are null), is often used to test the effects of many SNPs in GWAS studies (e.g. \citealp{price2006principal,gagnon2013removing,wang2017confounder}). 

\paragraph{The number of causes must go to infinity.}
The identification results in WB require \emph{consistency of substitute confounders} (Definition 4 of WB), which generally holds asymptotically as the number of causes, $m$, goes to infinity.  This is the case, for example, for probabilistic PCA and Poisson factorization, as discussed by WB and for which $(n+ m)log(nm)/(nm)\rightarrow0$ ensures consistency.  Consistency likely also requires either (a) a parametric factor model or (b) that a discrete variable with finite support suffices to control for confounding.
It is not immediately clear what estimands are defined and identified in this limit, since Theorems 6, 7, and 8 are written for finite $m$.  Furthermore, it is not clear whether identification holds 
for any finite $m$.
Of course, desirable frequentist properties for estimators of causal effects often require asymptotic arguments.  However, in most settings that argument is required for estimation but not for identification; here an asymptotic limit in both the number of causes and the number of subjects is required for unmeasured confounding to be controlled for and therefore for identification. 
 
However, the requirement that, in the limit, $Z$ be a deterministic function of $\mathbf{A}$ suggests that it cannot, in fact, control for confounding.  This is because such a $Z$ is independent of $Y$ given $\mathbf{A}$, which is not true of confounders (see the analysis of Lemma 4 above).  If causal effects are identifiable using such a $Z$, it must be because bias due to unmeasured confounding is estimable with a function of $\mathbf{A}$, and that function is not collinear with the causal effects themselves.  In this case the method would have to rely for identification not on ignorability, but rather on an assumption that a biased, confounded effect and its bias are simultaneously identified.

\section{Conclusion}

One of the most important roles of causal inference
in statistics and data science is to be transparent about the strong,
usually untestable assumptions under which causal inference is possible \citep{pearl2000causality, robins2001data}. The burden for transparency about assumptions is arguably greater in causal inference than in other areas of statistics, because it is crucial that scientists and consumers of research, e.g. policy makers or doctors, have the tools to reason about whether an association is in fact causal.
To that end, our best current understanding of when it is justified to use a substitute confounder based on a factor analysis 
to estimate causal estimands in the presence of unmeasured confounding is under these conditions/assumptions (some of which are explicit in WB):
\begin{enumerate}
\item No unmeasured single-cause confounders.
\item SUTVA
\item No M structures exist between A and Y.
\item The causes are not causally dependent.
\item No post-treatment variables are captured by Z.
\item Unmeasured multi-cause confounding is due to a dependence or clustering structure that is common to each cause and to the outcome.
\item Z is consistent, which may rule out confounding altogether (see discussion above).
\item In the limit as the number of causes and the number of observations go to infinity.

\item One of the following:
\begin{enumerate}
\item Confounding is due to a clustering indicator, treatment effects are constant in $Z$, continuous causes, and relative smoothness constraints on functions of the causes and of $Z$ identify joint treatment effects of all of the causes (WB Theorem 6).
\item Overlap for some causes identifies treatment effects for those causes (WB Theorem 7).  This is at best a semiparametric assumption given the definition of $Z$ in terms of the causes (see discussion of semiparametric and parametric assumptions above).
\item Common values of $Z$ identify conditional potential outcomes (WB Theorem 8).
\end{enumerate}
\end{enumerate}

Some of these assumptions may be able to be relaxed or replaced with different assumptions, but unfortunately -- we wish this were not the case! -- it is impossible to identify causal effects in the presence of unmeasured confounding with nonparametric or empirically verifiable assumptions.

\section{Response to WB}

We are grateful to WB for being willing to discuss their paper, and their rejoinder pointed out a few errors and points of clarification in an earlier version of our comment (noted by footnotes above). Nevertheless, we largely disagree with their rebuttals of our critiques of the deconfounder method. Most importantly:

\begin{itemize}
\item No formal argument is presented in either the paper or rejoinder to justify the claim that, because $Z$ is a function of many causes, it is precluded from inadvertently picking up single-cause colliders or mediators.   Let $C$ in Figure 1 (c) or $D$ in Figure 1 (b) be equal to a function $g(\cdot)$ of $A_m$ (and possibly a random error), and suppose that $Z$ is also a function of $g(A_m)$. Although this example is degenerate from a causal perspective, it shows that $Z$ can incorporate information about random variables that are downstream of a single cause.

\item Given that the definition of a multi-cause confounder in Appendix E (arXiv version; Appendix H in JASA) is about unconditional properties of a confounder $U$, this definition alone cannot rule out the possibility of $U$ being separable after conditioning on $Z$.

\item The claim that results in \cite{kallenberg2006foundations} license omitting the potential outcomes from the construction in equation (38) of WB (arXiv version; this is equation 37 in JASA) is incorrect.  This follows from the fact that potential outcomes are part of the relevant measurable space $S$ in Theorem 5.10 of Kallenberg.  Our point is that, if potential outcomes are not included in (38), then they cannot be included in (39) either (that is they cannot be included in equation 38 in JASA).

\item As a counterexample to the claim that differentiability of $f_1$ and $f_2$ suffices for Theorem 6, we note that $f_1$ could be equal to $f_2( f( \cdot ))$, in which case the two functions would be collinear.

 \end{itemize}
 
Finally, we wish to highlight the assumption that $Z$ is a deterministic function of the causes, which is crucial to the original paper and to WB's response. As  discussed above, this implies that $Z$ cannot be a confounder according to standard definitions of confounding (see Analysis of Lemma 4, above). In Lemma 4 WB argue that $Z$ can be treated as a pre-treatment random variable (confounders are pre-treatment by definition), but this is not true in general. A function's inputs are causally antecedent to its output, therefore a function of causes is a post-treatment variable.  Furthermore, since $Z$ is caused by each of the causes,  it is a collider between the causes and, in the absence of extra assumptions, conditioning on it induces dependence among the causes \citep{cole2009illustrating, elwert2014endogenous}. Even in a degenerate model in which a deterministic function $f(\mathbf{A})$ is equal to a pre-treatment variable, one cannot rule out that this same function $f(\mathbf{A})$, or some components of it, also operate as post-treatment random variables (see the first bullet point above).  

It may be feasible to construct limited models in which a post-treatment function of the causes is equal to a pre-treatment random variable, and therefore can be considered to be simultaneously pre-treatment and post-treatment, e.g. when $Z$ represents some types of clustering.  But a $Z$ that can be simultaneously pre- and post-treatment is pathological indeed; among other pathologies it is only possible in settings that violate the faithfulness assumption that underpins much of causal inference \citep{scheines1997introduction}.  Such a variable would not exist in most substantial settings in which we are interested in causal effects of multiple causes are of interest.

\section*{Acknowledgements}
We would like to thank Alex D'Amour, Susan Murphy, and Zach Wood-Doughty for helpful discussions.

\bibliographystyle{jasa}
\bibliography{refs}

Wang, Y. and Blei, D. "Blessings of Multiple Causes." \emph{Journal of the American Statistical Association, Theory and Methods}, (In press).
\end{document}